# Influence of AI in human lives

Meenu Varghese
*Dept. of Computing and Informatics*
*Bournemouth University*
Bournemouth, UK
s5434850@bournemouth.ac.uk

Satheesh Raj
*Dept. of Computing and Informatics*
*Bournemouth University*
Bournemouth, UK
s5426480@bournemouth.ac.uk

Vigneshwaran Venkatesh
*Dept. of Computing and Informatics*
*Bournemouth University*
Bournemouth, UK
s5431298@bournemouth.ac.uk

*Abstract-Artificial Intelligence is one of the most significant and prominent technological innovations which has reshaped all aspects of human life on the lines of ease from magnitudes like shopping, data collection, driving, everyday life, medical approach and many more. Contrastingly, although recent developments in both subjects that are backed by technology, progress on AI alongside CE must have mostly been undertaken in isolation, providing little understanding into how the two areas intersect. Artificial intelligence is now widely used in services, from back-office tasks to front-line interactions with customers.*

*This trend has accelerated in recent years. Artificial intelligence (AI)-based virtual assistants are changing successful engagement away from being dominated by humans and toward being dominated by technologies. As a result, people are expected to solve their own problems before calling customer care representatives, eventually emerging as a crucial component of providing services as value co-creators. AI-powered chats may potentially go awry, which could enrage, perplex, and anger customers. Considering all these, the main objectives of this study will engage the following*

- *To identify the alterations in the scope of human searches for information offered by the application of AI?*
- *To analyse how AI helps in the way someone drives the car*
- *To evaluate how AI has changed the way customer interact with the customers*

## I. INTRODUCTION

Artificial intelligence is the imitation of superior intelligence capabilities by technology, specifically Computer operating systems. Instances of specialised AI technologies include experience and understanding solutions, engineering, natural language processing, and analytical approaches. As even the hysteria surrounding AI grows, businesses have rushed to highlight regardless of whether their product offerings incorporate AI. Often, everything people mean by AI is merely one of its components, such as machine learning. Artificial intelligence (AI) necessitates a framework of sophisticated operating systems for the development and integration of machine learning methodologies. Automated systems often consume a sizable quantity of training data, which is then analysed for commonalities and associations before using comparable frameworks to anticipate potential consequences. A facial recognition software system potentially learn to recognise faces and other objects in photos by examining countless examples, just like a chatbot could learn to have natural discussions with friends by analysing examples of text messages.

Basically there are three types of cognitive skills in the daily lives of human being on which the main focus is given by the application of AI.

- *Learning Process*: This branch of AI research focuses on gathering information and creating the systems that will allow the facts to be turned into intelligence. The rules, usually referred to as protocols, give computing technology specific instructions on how to do a particular action.
- *Reasoning Process*: The optimum procedure to obtain a specific outcome is a consideration in some other area of AI technology.
- *Self-correction Process*: Perpetual methodology improvement is a feature of AI development that helps make sure it produces results that are considerably increasingly accurate.

Numerous innovative technologies, from corporate applications to embedded devices, are connected to the

World Wide Web due to AI applications. Technology, communications, applications, data/cloud infrastructure, and several fresh company operations are just a few of the domains and sectors of the economy that artificial intelligence can be used in. Essentially said, the AI can indeed be divided into domestic AI and corporate AI, both of which execute various implementation duties despite sharing comparable characteristics and activities. In reality, there are still a lot of obstacles in AI that need to be resolved, including economic systems, privacy and security concerns, telecommunications, and power consumption.

The broader public and mainstream media both been particularly interested in contemporary advancements in artificial intelligence. Considering the moral implications of Artificial Intelligence systems like robotics, chatbots, simulations, as well as other virtual assistants is a key area of research and development because these technologies shift as being seen as tools to intelligent entities and employees. The degree to which civilization and technological mechanisms are able to address these problems will play a role in determining human trust level towards AI, as well as its eventual societal impact and continuance.

The main research questions that will be addressed in the study are that of

- What alterations in the scope of human searches for information are offered by the application of AI?
- How AI helps in the way someone drives the car?
- How AI has changed the way customer interact with the customers?
- What modifications have been simulated by AI in the field of medical science?
- What kind of developments are made by AI in the realm of human shopping?
- What positive and negative influence of AI is casted on the normal human life due to everyday applications?
- How tasks are automated using the technological advancements of AI?

Contradictory to the ominous depictions of something like a dystopian nightmare in the television and also in contemporary literature, in which AI applications rule the globe and therefore are essentially concerned with fomenting rebellion, AI has already been altering human everyday lives, primarily in addition to enhancing population wellbeing, protection, and performance. The goal of artificial intelligence is to promote inclusive development as well as contentment in a better environment by facilitating the implementation of expert machines in accordance with natural constitutional values and principles.

## II. ALTERNATIONS OFFERED BY AI IN THE WAY HUMAN SEARCHES FOR INFORMATION

The domain of smartphone technology is also moving toward employing machine intelligence as a result of the swift advancement of AI capabilities. The utilization of conversational user interface design is growing. Voice-activated operating systems, notably those used by modern smart voice assistants like Alexa, Bixby, as well as Google Assistant, are among the most popular that is being used today. According to numerous schooling specialists, the educational system needs electronic contributions in the current environment [22]. Technological would increase accessibility and inclusivity in learning. Introducing AI to the actual class environment can achieve this goal. Every industry, including the business and healthcare systems, has already seen AI's influence. The speech synthesis systems Cortana by Windows, Siri by Apple, and Alexa by Amazon are considered to be good instances of AI technology and therefore can imitate intellectual ability. Several tools help people learn new things and improve the ability to make decisions. This method, sometimes referred to as machine learning, may tackle numerous problems in the field of learning with ease.

Artificial intelligence's advent has allowed online learning becoming individualised for each individual. It cuts down the amount of time, money, and aggravation associated with finishing internet or android programs. Communication lessons now include a lot of discussion-focused sessions attributable to AI. It helps learners and teachers both preserve money [6]. The learners will also have the chance to pick up the language quickly because the English teaching assistant can create many sessions that are activity-oriented. Instead of using a one-size-fits-all approach to teaching and learning, AI may be used to create individualised educational experiences that let students focus on their own weak spots and gain from customised comments. AI-driven English language learning is so geared [46].

A person typically will have to endure until their opportunity to speak up and receive comments. The amount of engagement between one teacher and a student is constrained. In this situation, the chatbot in English class aids in personal practise. AI encourages gamification of both the language-learning procedure. Numerous chat programmes have entertaining features, and students who utilise them and achieve their conversational objectives may receive some kind of reward [28]. The chat robot may react continually and provides an immediate reaction to any question whether people feel like speaking in English, even during the early hours of the morning. As robots lack human behavioural traits like glazed eyes and raising heads, individuals cannot be judged in this language learning and feel ashamed. Although if students are unable to understand a word correctly, they do not feel embarrassed because machines do not pass judgement. Whenever people erroneously make a mistake, they do not even frown or raise their heads. Whenever there is no judgement or shame, learners can openly discuss and practise.

Two distinct kinds of robotics can be utilized to communicate and practise the English language. Firstly, chatbots that focus on language development are developed to instruct non-native speakers, as well as secondly, chatbots that mimic the intonation of native speakers assist students in learning English in context. These days, chatbots are performing a lot of duties. These AI-enabled chatbots evaluate language learners' English language development skills. It helps students to perform drills and engage in greater conversation [7]. It offers a structured learning journey so that students can progress at their individual rate and according to their personal inclinations. Certain applications, like Mondly, an online reading comprehension platform with a smartphone app, can enhance the acquisition procedure for English. Its programmes support chatbots and can be utilized on both desktop and mobile devices. The creation of the mobile applications ELSA (English Language Speech Assistant), which helps with articulation, was also greatly aided by AI [31]. It offers a variety of educational possibilities, including leisure or management can be characterized, and it includes a list of vocabulary items and process is the identification. Not only does it identify pronunciation errors, but it can also demonstrate ways to do so.

### III. HOW AI HELPS IN THE WAY SOMEONE DRIVES THE CAR

Today's culture is undergoing a change due to artificial intelligence. AI technologies for automated vehicles are continuously being promoted by scientists and developers in the automobile industry. AI network should first satisfy a rigorous operational safety inspection before it can be used in cars in large quantities. In current history, a multitude of self-driving automobile demonstration projects have been displayed [16]. Several pilot projects all share the characteristic that AI based methods are used to perform some driving functions, including trajectory tracking, environmental sustainability, and even steering wheel manipulation. The automobile sector is increasingly shifting its attention away from developing and presenting experimental automobiles and toward serial manufacturing due to the program or intervention of the AI-driven independent concepts [24]. The main difficulty now is to integrate AI into mass-produced cars while maintaining regulatory standards.

Automated vehicles with self-driving capabilities are indeed a good example of intelligent innovation. The mental interactions, including such goal determination, equipment creation, administration utilization, and their interactions with the environment, govern automated automobiles. Standalone is the interface that enables the vehicle activities to be performed [36]. The main objective of the creation of autonomous vehicles is self-driving cars. The effectiveness and convincing in the conveyance that is dependent on the automotives can be modified and improved by the driverless cars and automobiles. Considering its advantages, it needs to be properly routed, made decisions, moved, and controlled on a common roadway with passengers as well as other automobiles that are either human-centered, self-driving person aided automobiles, or completely self-driving cars.

Both in academic and business communities, the creation of self-driving vehicles is a rapidly growing sector. It is a part of a bigger discipline that also deals with developing technologies using ideas from automated systems and artificial intelligence. Numerous challenges need to be overcome in order to provide self-driving cars in a manner that is simply understood by customers. Research shows that while many people are at ease with the concept of having AI manage their homes or gatherings, just under half of them are equally at peace with the concept of having AI manage their cars. Insurance companies are worried about concerns with insurance coverage and how they can prove that something or an individual provoked a catastrophe at the same time. Humans think that self-driving automobiles that operate transparently will win over more consumers' confidence, which is essential for adoption and therefore will open the door to their widespread use and monetization [38]. In order to create increasingly reliable self-driving cars, researchers explore throughout this study how to approach the articulation and modelling of visibility as both a Non-Functional Requirement (NFR).

Self-driving automobiles are undoubtedly a number of the most frequently debated and investigated subjects in recent decades. Considering the assumption that people place themselves in a particular area of the automobile sector, according to all criteria these technologies, as a subsequent robotic transformation, belonged to the industrial robotics [39]. Incorporating the larger subject of automation, such as environmental sensing, decision - making process, and management, reproducing the challenging task of human driven cars by an inertial navigation system presents innumerable technological hurdles.

In order to provide self-driving automobiles, the automobile manufacturers are adopting new difficulties, which in turn calls for ever-more-complex systems and equipment. AI specifically is being used by computer programmers to implement the tools necessary for a driverless vehicle to function [13]. Due to this, automobile electronics are now more diverse and feature multi-core engines and specialised co-processors that can handle the sophisticated software needed for machine learning and artificial intelligence. Common processes employed in AI as well as machine learning algorithms can be greatly accelerated by these hardware resources.

## IV. AI IS CHANGING THE WAY CUSTOMER INTERACT WITH SERVICES

AI is expected to have fundamental structural effects on consultancy, which will change how companies communicate with prospective consumers. Customer experience, which simply gauges how customers to communicate with businesses, goods, and businesses over the previous 10 years, has gone to the top of many senior management organisational wish lists. Contrarily, development on AI with CE has, up to now, largely been conducted in isolation, giving little insight into how the two domains interact, despite recent advances in both disciplines that are supported by material [1]. The utilization of Artificial Intelligence in services has increased dramatically in recent years, with wide range of methods from back-office operations to front-line customer care engagements. The fourth major manufacturing period, known as Industry 4.0, is defined by a convergence of technologies that dissolves the boundaries between both the mechanical, technological, and physiological domains. AI-based service management is a component of this time in history. Although company uniformity has always been associated with automation, today's automated service engagements deliver progressively individualised service while lowering unpredictability and mistake.

Automation, nanoelectronics, cognitive computing, nanotechnology, the Internet of Things/Everything, as well as 5G are just a handful of the advancements in technology that form the foundation of Industry 4.0 and therefore will surely have a consequence on how clients engage with services [14]. For instance, KLM's Spencer this same chatbot provides customer relations, McDonald's terminals streamline the delivery of services, and Royal Caribbean's Quantum of the Seas has automated waiters. Additionally, since COVID-19 began, the need for driverless food delivery robots has skyrocketed. A key component of Industry 4.0 is the use of enormous volumes of Big Data to develop AI systems. Statistical intelligence, which could be used to better explain customer satisfaction and other service-related problems, might span between structured through unstructured data that details standalone treatment. AI uses huge data to identify trends in past behaviour, provide increasingly precise predictions for future behaviour, and generate necessary suggestions [30]. For instance, Google Ads analyses and offers certain advertising that this really believes consumers would find most intriguing by monitoring users' surfing and purchasing trends.

It is not really remarkable that computerized encounters are thought to having tremendous potential for consumer experience. For example, according to Forbes, by 2025, 95% of business and customer contacts will be handled entirely by digital technology. Thus, it is anticipated that AI-enabled customer experience would expand quickly in the generations to follow and could probably replace some social communication, such as automated supermarkets checkout lines, which may have been highly desired inside a (post-) COVID world [5]. Virtual assistants based on AI are transforming effective engagement from being dominated by humans to becoming dominated by technologies. Individuals are consequently required to address problems on their own before contacting customer service agents, eventually emerging as a key component of delivering services as value co-creators [45]. AI-powered conversations may also go wrong, which could cause consumer annoyance, bewilderment, and fury. In order to increase organizational excellence and improve the client encounter, AI is widely used in service - based organizations. The manner consumers interact with companies is being revolutionised by artificial intelligence. Customer interactions utilising AI are not well studied empirically.

Artificial intelligence seems to have the ability to completely change how companies communicate with their clients. AI is distinct from natural intelligence because it relies heavily on quick data computation [44]. The capability to analyze and synthesize data into knowledge to guide goal-directed behaviours is a broad definition of cognition in AI. AI is more clearly defined as "supercomputing agents that perform actions sensibly," which are created to mimic the human intelligence and outperforming it in terms of precision. Organizations have embraced AI technology backed by business intelligence more and more as a solution to ongoing profitability challenges, compressed marketing timelines, and higher expectations from customers. Improved client partnerships could result from changing how businesses communicate with their clients [34]. Particularly, improvements in AI have the potential to enhance the customer engagement by enhancing businesses' understanding of those clients' inclinations and buying habits. Consequently, appropriately implementing AI technology at several important consumer multiple touchpoints can result in considerable advantages for businesses as well as a potential rise in client satisfaction.

## V. MODIFICATIONS DONE BY AI IN THE FIELD OF MEDICAL TREATMENTS

Healthcare provider is changing as a result of the development of new machine learning and artificial intelligence approaches. Contemporary AI-based techniques have also already improved the accuracy and efficacy of diagnosis and therapy across a wide range of disciplines, especially when paired with the enormous strides in performing calculations [15]. Due to the increasing attention AI is getting in the domain of radiography, many experts have predicted that it may potentially even replace doctors. These concepts raise the question of whether physicians in particular specializations will ever be replaced by AI-based technology, or if their contemporary jobs will be improved.

AI demonstrates a broad range of production capabilities for unique prostheses and cutting-edge hospital instruments. It delivers a kind of electronic healthcare and a full tracking system which optimally balance costs and time while meeting the specific needs of the patient/medical sector. Incorporating technology, intelligent devices, and technical intelligence is an imaginative way to explore new ideas and advance the healthcare industry [18]. Industry 4.0 produces controlled, significant healthcare gadgets that are extensively developed to suit clinical preferences. In the healthcare industry, technological transformation promotes mechanization and opens up new production options. It develops a new fictional environment with the assistance of something like the Internet of things as well as the Internet of services [40]. With both the aid of modern fabrication techniques, programming, scanners, robotics, and some other cutting-edge communication technology, it fosters communication and data sharing. In the future, AI may open up fresh possibilities and suggest novel approaches to medical care.

Apparently, a computer programme with artificial intelligence can identify skin cancer with greater precision than a dermatologist with board certification. Even better, the algorithm can complete it more quickly and effectively, assuming only a training set of information as opposed to 10 years of costly and time-consuming medical school [4]. Although it could seem that all of this innovation will eventually replace doctors, a thorough examination of the potential applications for that in the provision of health services is necessary to understand its existing advantages, drawbacks, and ethical concerns.

Just about every area of pharmacy can benefit from the use of artificial intelligence, including encompasses the domains of cognitive computing, speech recognition, and automation. The long term benefits for biomedicine, professional examinations, as well as the provision of medicine seem unlimited [32]. Due to its incredible capacity to integrate and comprehend vast amounts of medical research, artificial intelligence potentially plays an important role in diagnosing, medical decision-making, and personalized medication. For examples, AI-based mammograms diagnostic tools are assisting in the detection of colon cancer and giving medical professionals a "accurate diagnosis" in the procedure. The ability of advanced computerised character subjects to conduct meaningful conversations has some more implications for the diagnosis and treatment of psychiatric problems. Cases of AI applications which also extend into the material realm include robotic replacements, metabolic activity supportive services, and transportable exploiters that assist in the management of healthcare.

However, as AI systems have the potential to seriously jeopardise individual needs, welfare, and confidentiality, a different generation of moral problems are presented by this potent innovation that have to be recognised and addressed. Furthermore, present AI legislation and professional ethics fall the behind advancements AI has accomplished in the healthcare industry. The medical establishment is still unaware of the ethical challenges that developing Artificial intelligence can pose, despite some efforts to participate in these ethical discussions emerging [26]. As a result, there seems to be a probability that using AI in pharmaceutical research will be practical. Fewer of the older medications, if they are proved to be effective towards SARS-CoV-2, can be quickly used to treatment COVID-19 sufferers due to prior optimal experience in individuals. AI and medicine working together can dramatically increase the effectiveness of therapeutic applications.

Among the most significant current revelations in healthcare is method measures, that has the ability to transform the conventional symptom-driven clinical practice by enabling previous treatments utilising cutting-edge diagnostic information and crafting improved and more affordable individualised treatments [23]. The ability to make exhaustive patient records in addition to more general issues to supervise as well as differentiate among both ill but instead generally healthy people will result in a better knowledge of bio-indicators that really can indicate improvements in health, which also will result in the determination of the correct route to personalised and birth rate prescription medication.

## VI. DEVELOPMENTS MADE BY AI IN THE WAY PEOPLE SHOP

Over through the past ten years, the e-commerce sector has grown quickly. Understanding digital clients' wants and behaviour patterns is essential given their increasing rise. E-commerce companies adopted AI-enabled technologies as a result to determine client desires and preferences for digital merchandise and services. The artificial intelligence keeps track of the users ' preferences, buying habits, spending patterns, and typical dollar amount during a certain time frame. It offers E-commerce Businesses comprehensive personal data [19]. As a result, this data allows companies to customise their goods and services to meet the unique preferences and demands of their clients.

The AI helps the consumer choose and purchase the most appropriate recommendation items / solutions by offering a variety of tactics for specific suggestions, reductions, and numerous offers. The pandemic altered customer behaviour, particularly with regard to non-essential products, wherein profit margin fell sharply in the third and second quarters of 2020. Although isolation and loneliness remained the most effective way to lessen health issues related to the infection, many customers began adopting internet shopping to the disadvantage of brick-and-mortar establishments [20]. Companies are turning to e-commerce in this environment of profound and rapid change in an effort to meet peoples' evolving wants.

The simplicity of being used, the platform's or application's appearance, and the public's confidence in the handling of the confidential information taken are the three most crucial factors for customers, according to studies who have noted that internet purchases has drawn a lot of interest in recent years.

Since this industry produces a large amount of information that can be used by this data-driven innovation, the development of AI based technologies is quickly changing the sector by improving the complete experience. Several studies have looked into the real-world applications of AI-based technologies in retail and wholesale over the ages [21]. The most fundamental need for humans are genetically determined. One way to meet metabolic requirements for things like apparel, nourishment, and hydration is to go shopping. The bulk of citizens love to purchase in shopping malls since they are more comfortable and safer and provide high visitor movement because of their consistent spatial arrangement.

Minimal emphasis has been placed on applying AI to improve the in-store experiences through smartphone platforms, particularly in today's environment where clients generally avoid pointless social connections. When opposed to internet buying done through web platforms, mobile commerce apps give businesses greater power to boost participation and make use of numerous AI technologies that may be incorporated into the apps [29]. Businesses from all over the world began implementing AI technologies to help this industry grow. Only with application of AI across several facets of retail, including warehousing and distribution that really can move objects and get items ready for shipping, Amazon was among the most prominent innovators in digitizing this field.

Businesses are getting increasingly interested in chatbots, sometimes known as virtual assistants, primarily to their application since this technology promises to enhance communication, streamline communication, and offer a customised customer experience. Applications can make utilisation ML methods, computational linguistics, and naturalistic computer vision. It has become clear that each person have the capability to play a crucial role in the elevation of society and the neighbourhood as a result of the advancement of humankind and technology [1]. Individuals who are less fortunate compared to others would require greater fortitude and tenacity to rise above their present circumstances and get over common and unavoidable challenges in order to mimic an atmosphere where production and innovation abound.

## VII. POSITIVE AND NEGATIVE INFLUENCE OF AI IN HUMAN LIFE

AI technology has been applied recently to produce marketing communications. The influence of AI on civilization has already been profound. They determine whether there is major controversy about what comprises "ethical AI" but whether those ideas are disparate, with such a set of mutually agreed-upon criteria. The creation and application of AI have the potential to have both beneficial and bad consequences for society, to lessen or accentuate current disparities, to solve current issues or create new ones [37]. A review of AI's influence on achieving the Sustainable Development Goals is necessary given its existence and steadily expanding influence on numerous areas. A wider spectrum of industries are being shaped by AI's rise. For contrast, both immediately and later on, AI is anticipated to have an impact on worldwide production, diversity and inclusion, sustainability results, and numerous other areas. Two very different positive and negative effects on environmental sustainability are indicated by the anticipated possible effects of AI.

While this research characterised environmental sustainability as the 17 Sustainable Development Goals as well as 169 targets globally consented on in the 2030 Sustainable development Agenda, to deadline there has been no piece of research methodically evaluating the magnitude toward which AI could perhaps influence all facets of sustainability [17]. AI systems have enormous ability to boost not simply the promotion of tourism-related services and products in addition to the influence of sustainable transportation practises. The acceptance of AI by customers will determine how well initiatives to promote good reinforcement will work. AI is increasingly ingrained in the daily lives, both at home and at work. AI-based virtual platforms, which seem to be increasingly widely accessible and have a broad spectrum of applications possibilities, are an important field of development.

Technology tools powered by AI offer a lot of advantages, but they could also pose risks. On only one extreme, they are anticipated to replace people in menial activities and remove unnecessary resources and time for even more difficult ones. For examples, IBM claims that chatbots can aid in shaving 30% off the cost of customer care [2]. AI is quickly paving the way for new developments in the realms of commerce, commercial strategy, and public administration. Deep learning-capable robotics and automated systems have had a profoundly disruptive and empowering effect on the industry, governance, and civilization. Companies are also having an impact on more general trends in international environmentalism.

Whenever AI affects the society, it may foreshadow a socialist utopia in which both people and machines live in peace or it may foreshadow a nightmarish future marked by war, famine, and pain. Scientists and practitioners have given AI advancements a great deal of attention, and this has led to a diverse variety of advantageous potential for AI use in the government service [11]. In light of something like this, there has been a growing need for a comprehensive knowledge of the breadth and effect of AI-based technologies and related difficulties. AI is becoming a more important player in corporate strategy as a result of the rising quantity of information and knowledge and also the limiting computing power of the human brain.

Business depends on management who are prepared to delegate responsibility and oversight to AI-based processes that result in order to take advantage of the possibilities that AI offers for strategic choices. This contributes to a better understanding of social cognition in choosing assignment to AI in the setting of strategy formulation by examining the propensity to outsource business choices and behavioural responses to results of transferred judgments [8]. Convolutional AI are already frequently utilised in mission-critical applications that influence personal lives, such as hospitals, self-driving cars, and the military. AI networks' usage in mission-critical operations is, nevertheless, constrained by their black-box character, which raises moral and legal issues that erode confidence [12]. An area of AI known as explainable artificial intelligence (XAI) aims to develop instruments, methodologies, and programs which may produce interpretable, logical, and human-understandable interpretations of AI judgments that possess a high calibre.

## VIII. AUTOMATION OF TASK USING THE PLATFORMS OF AI

As AI is used increasingly frequently in modern civilisation, it is necessary to examine how people and AI interact. The technological advancement has progressed beyond the simple programming of subjective interaction into a mechanism to the creation of computers that "know how" to automatically collect the data they require, understand from all of this, and operate in their environments. Interestingly, this requirement has theoretical roots that date back to the creation of machines, therefore it is not a recent one [25]. A leading scientific problem to find and create substances for new and upcoming innovations is to speed up substances development by combining mechanization and artificial intelligence.

Although the semiconductor devices components scientific community has successfully used a wide range of high throughput technologies and analytical strategies to speed up specific research activities, a really transformative advancement of commodities development is still not yet entirely realised. For many years, mechanization has supported human workers [3]. The primary benefactors of mechanization or the tasks that enabled it has always been manufacturing and transportation. Nowadays, many tasks that were originally performed by human beings are automated using AI and machine learning. Technology as a consequence helps humans in very many facets of ordinary routine. Completely automated database design and sophisticated analytics tools help CEOs with elevated managerial activities, physicians with diagnostics, and magistrates in legal proceedings, among other professionals.

Human resource administration analysis and pragmatic developments now point to a day in the future when administrators will work with intelligent automation to carry out tasks like scheduling, hiring employees, and maintenance. Nevertheless, the effects of mechanization go beyond issues with consumerism or efficiency/effectiveness [33]. This one has been discovered that the adoption of automation in traditional industries does have an effect on job responsibilities, inspiration, and general well-being at employment. Modernization of management labour, in people's opinion, may have a similar effect. The capability of recruitment administrators to perform their duties already has been significantly impacted by automation, notably in the area of personnel selection as a dedicated management duty.

For instance, they might well have technologies that gather data from applications and offer applicant scores based on computer-assisted evaluation of applications or interview skills. On the one extreme, this might boost efficiency, providing employees additional time to concentrate on other tasks [43]. Automatic recruitment processes have an impact on how hiring managers process the information. There are four primary types of automated functions on the lines of information gathering, screening and assessment, recommendation of solutions, and movement implementation. As greater numbers of these activities are linked into networks, the amount of automation will increase [9]. It is indeed conceivable because collecting data will involve video interpretation that is computerized. It really is conceivable that underlining specific words within those translations will facilitate data processing and assessment.

Technologies that provide judgement recommendations learnt from statistical information may be used to complete multiclass classification (such as sorting appropriate from unsuitable applicants) or forecasting activities. When these actions have yielded outcomes, individuals may be presented information to aid in decision-making [42].

They are therefore helpful to recruiting firms in a range of activities like gathering and evaluating information, accepting or rejecting candidates, and whose conclusions could serve as an additional online resource for judgement call. Algorithms that, for instance, reduce the need to review primary user information may lead some to believe that its significant contribution on the recruiting process is lessened. The terms of projects are considered in terms of autonomy and accountability may have repercussions that could reduce satisfaction with one's job [27]. The application of AI and ML to support organizational tasks and functions has gained prominence. Contrary to more traditional uses such as manufacture or transportation, the deployment of mechanization in administration is not comprehended.

Automated processes, which can be conceived of as progressively complicated and advanced means of automatically getting or integrating data, have been made possible by artificial intelligence and machine learning. In particularly, several cutting-edge approaches use webcams as well as mics as instruments to collect data from participants, natural language recognition to gather data from application program for process control, and machine learning algorithms to judge the aptitude of candidates [10]. There are numerous approaches to integrate computerized signal collection and judgement technologies for selection of employees. In addition to the specific duties delegated to an automated service, it is crucial to take into account the timeliness of when to provide recruiting supervisors with both the findings of a computerized method.

For instance, computerized support prior to the processing, including a catalog of automated assessment and employment services and products may be offered by computerized personnel methodologies. Generally speaking, the solution provides after evaluating the candidate's data. These findings and further information about the candidates are provided to recruitment supervisors. As a consequence, they possess accessibility to the results of the automated process and can look up more user information [35]. The propensity for centering consequences, particularly concentrate hiring managers' attention only on applicants with the highest marks, is a significant downside. Evidence from traditional cloud computing fields also demonstrates that people immediately view comparable technologies as incredibly reliable that could lead decision-makers to embrace ideas without carefully considering other, potentially conflicting reports or the suitability of candidates [41].

# IX CONCLUSION

The above data provides ample amount of factual data and informative insights on the subject matter projecting the influence of AI in human lives from which it can be inferred that Artificial intelligence is when equipment, primarily computerized software platforms, imitates the characteristics of intelligence. Examples of specialist AI technologies include construction, computational linguistics and analytical methods. Corporations have hurried to emphasise if AI is present in their product portfolio as the excitement surrounding AI rises. Technological solutions frequently take in a substantial number of training examples, which is then examined for patterns and relationships before employing analogous concepts to forecast probable outcomes.

Similar to how a chatbot could pick up on genuine conversations between people by studying text message samples, a facial recognition program system could theoretically learn to identify individuals and other things in images by studying many examples. Technology would improve learning's affordability and diversity. This objective might be accomplished by integrating AI into the classroom setting. Every sector of society, particularly business and medicine institutions, has already felt the impact of AI. As good examples of Artificial intelligence, the continuous speech programmes Cortana by Windows, Siri by Apple, and Alexa by Amazon can mimic human intelligence. Many tools aid in knowledge acquisition and decision-making improvement.

AI promotes the personalization of both language acquisition and learning in general. Many chat programmes offer amusing features, and students who use them and succeed in their conversational goals could be rewarded. Even in the late morning, the chat robot can respond instantly and repeatedly to inquiries about whether people feel like communicating in English. It enables them to carry out exercises and have more meaningful conversations. It provides a planned classroom journey so that learners can advance at their own pace and in accordance with their particular preferences. A few tools, like Mondly, an online tool for comprehension skills with a mobile app, can speed up the process of learning English.

Its applications are accessible on desktop and mobile devices and support chatbots. The application of AI-based technologies to carry out specific driving tasks, including as trajectories monitoring, environmental sustainability, and even steering wheel manipulation, is a common theme among several pilot projects. Due to the programme or participation of the AI-driven independent concepts, the automotive industry is rapidly turning its focus away from developing and displaying experimental autos and toward serial manufacturing. In order to supply self-driving automobiles in a way that is easily comprehended by customers, a number of difficulties must be overcome. According to research, while many individuals are comfortable with the idea of having AI handle their homes or meetings, just about half are as comfortable with the idea of having AI run their vehicles.